\documentclass[epjST]{svjour}
\usepackage{graphics}
\begin{document}
\title{Noise Can Reduce Disorder in Chaotic Dynamics}
\author{Denis S.\ Goldobin\inst{1}\fnmsep\inst{2}\fnmsep\thanks{\email{Denis.Goldobin@gmail.com}}}
\institute{Institute of Continuous Media Mechanics, UB RAS,
             Perm 614013, Russia\and
           Department of Mathematics, University of Leicester,
             Leicester LE1 7RH, UK, EU
           }

\abstract{
We evoke the idea of representation of the chaotic attractor by
the set of unstable periodic orbits and disclose a novel
noise-induced ordering phenomenon. For long unstable periodic
orbits forming the strange attractor the weights (or natural
measure) is generally highly inhomogeneous over the set, either
diminishing or enhancing the contribution of these orbits into
system dynamics. We show analytically and numerically a weak noise
to reduce this inhomogeneity and, additionally to obvious
perturbing impact, make a regularizing influence on the chaotic
dynamics. This universal effect is rooted into the nature of
deterministic chaos.
 }


\maketitle

\section{Introduction}
During few recent decades, noise was found to have ability not
only to make an obvious disordering impact, but also to play a
constructive role, increasing degree of order in system
dynamics~\cite{McNamara-Wiesenfeld-1989,Gammaitoni-etal-1998,Pikovsky-Kurths-1997,Anishchenko-1995,Zhou_etal-2002,Teramae-Tanaka-2004,Goldobin-Pikovsky-2005a,Goldobin-Pikovsky-2005b,Goldobin-etal-2010,Anderson-1958,Maynard-2001,Goldobin-Shklyaeva-2009,Goldobin-Shklyaeva-2013,Altmann-Endler-2010}.
Most remarkable advances are the phenomena of
stochastic~\cite{McNamara-Wiesenfeld-1989,Gammaitoni-etal-1998}
and coherence resonances~\cite{Pikovsky-Kurths-1997}; the
suppression of deterministic chaos in dynamics of dissipative
systems by noise (thoroughly discussed, {\it e.g.},
in~\cite{Anishchenko-1995}); noise-induced enhancement of the
phase synchronization of chaotic systems~\cite{Zhou_etal-2002};
synchronization by common
noise~\cite{Teramae-Tanaka-2004,Goldobin-Pikovsky-2005a,Goldobin-Pikovsky-2005b,Goldobin-etal-2010};
Anderson localization~\cite{Anderson-1958,Maynard-2001}; the
analog of the latter phenomenon for dissipative
systems~\cite{Goldobin-Shklyaeva-2009,Goldobin-Shklyaeva-2013};
{\it etc}. In this paper we disclose a novel phenomenon, where
{\it weak} noise reduces disorder in chaotic dynamics. The effect
is rooted in inhomogeneity of the distribution of natural measure
over the strange attractor and impact of noise on this
distribution.

First, we demonstrate this inhomogeneity and reason for the
consequences expected from it. Then we construct an analytical
theory for the noise impact. Finally, the predictions of the
analytical theory are underpinned with results of simulation for
paradigmatic chaotic system, the Lorenz one~\cite{Lorenz-1963},
where one can see the disorder reduction effect with divers
quantifiers of regularity, and for the mapping which is relevant
for one of the classical routes of the transition from regular
dynamics to chaotic one~\cite{Lyubimov-Zaks-1983}.

In our treatment we relay on the idea of representation of chaotic
attractor by the set of UPOs embedded into it
(\cite{Grebogi-Ott-Yorke-1988,Ott_book}, {\it etc.}). On this way,
for instance, one can evaluate the average $\overline{A}$ for the
chaotic regime as
 $\overline{A}=\int A({\bf x})\mu({\bf x})d{\bf x}$,
where $\mu$ is the natural measure, and integration is performed
over the whole attractor. It was shown
in~\cite{Grebogi-Ott-Yorke-1988} (see also the text
book~\cite{Ott_book}) that the distribution of natural measure
over attractor can be recovered from properties of the UPOs. For
two-dimensional maps (and therefore three-dimensional dynamic
systems where one can use Poincar\'e section to construct such a
map) the natural measure of an UPO is $\mu\propto\Lambda^{-1}$
where $\Lambda$ is the largest multiplier of perturbations of this
UPO. This result is treated to be valid in higher dimensions as
long as we have only one positive Lyapunov
exponent~\cite{Grebogi-Ott-Yorke-1988}. The approach was employed
already in~\cite{Eckhardt-Ott-1994} for evaluation of fractal
characteristics of the Lorenz attractor from the UPOs data
weighted with measure $\mu\propto\Lambda^{-1}$.

The effect under consideration may shed light on fine mechanisms
(or one of them) of phenomena similar to the noise-induced
enhancement of the phase synchronization of chaotic
systems~\cite{Zhou_etal-2002}.

\begin{figure}[!t]
\center{
\resizebox{0.60\columnwidth}{!}{%
 \includegraphics
 {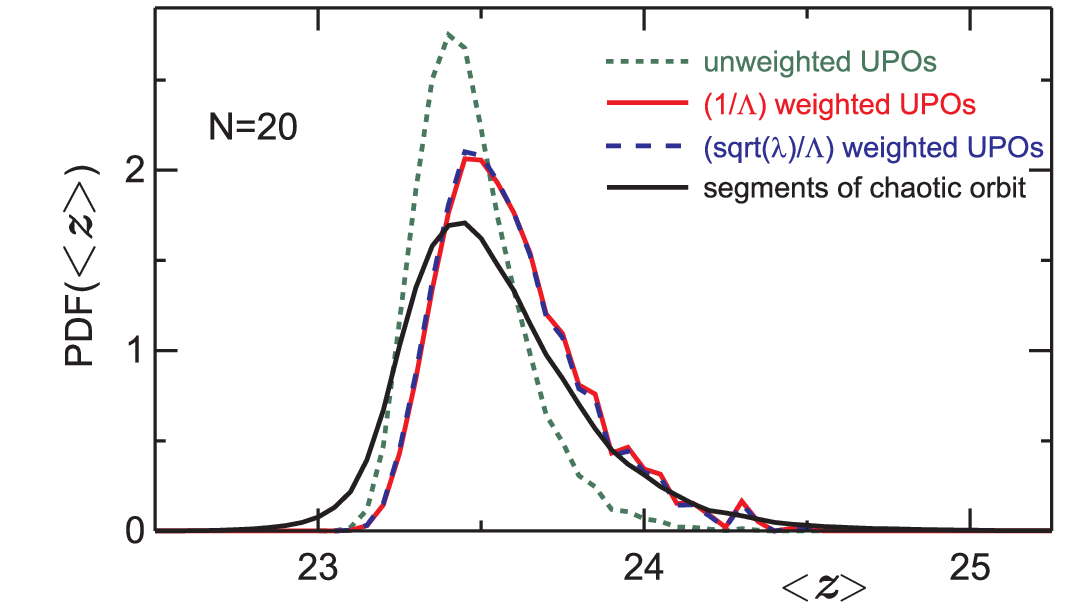} }
}
  \caption{Probability density function
of $\langle{z}\rangle$ calculated for the Lorenz system with
classical set of parameters ($\sigma=10$, $b=8/3$, $r=28$) over
UPOs of length $N=20$ without weight (doted green line); with
weight $\Lambda^{-1}$ (solid red line), which corresponds to UPO's
measure within chaotic attractor in noise-free case; with weight
$\sqrt{\lambda}/\Lambda$ (dashed blue line), which corresponds to
the maximal correction of the weight induced by a weak noise; and
$20$-loop segments of chaotic orbit (solid black line).
}
  \label{fig1}
\end{figure}

\section{Inhomogeneity of natural measure distribution}
We consider the Lorenz system~\cite{Lorenz-1963}, as an example,
\begin{equation}
\frac{dx}{dt}=\sigma(y-x),\quad
\frac{dy}{dt}=rx-y-xz,\quad
\frac{dz}{dt}=xy-bz.
\label{eq01}
\end{equation}
Recently~\cite{Zaks-Goldobin-2010}, we reported the results of
calculation of all UPOs with length $N\le20$ (by ``lenght'' of UPO
we mean the number of loops).  In Fig.\ref{fig1} one can see
probability density function (PDF) of $\langle{z}\rangle$
calculated over UPOs of length $N=20$, the distribution weighted
with $\Lambda^{-1}$ and PDF of $\langle{z}\rangle$ calculated over
$20$-loop segments of the chaotic trajectory.  The unweighted and
weighted distributions are remarkably dissimilar and the latter
matches the chaotic averaging much better. The weighted
distribution possesses a relatively big tail for large
$\langle{z}\rangle$, where there is a quite few number of UPOs
(unweighted PDF nearly riches zero next to
$\langle{z}\rangle=24$).  The origin of this difference is a huge
dispersion of multipliers for long UPOs: for $N=20$ the smallest
multiplier $\Lambda=62\,200$ is by factor $72.8$ smaller than the
largest one $\Lambda=4\,527\,160$. Noteworthy, the weighted
distribution is significantly broader than the unweighted one.
This feature implies a large contribution of relatively small
number of UPOs with extreme deviation of properties from the
typical ones ({\it e.g.}, in Fig.\,\ref{fig1} these UPOs have
untypically large values of $\langle{z}\rangle$). In terms of
temporal evolution, the system dynamics is strongly inhomogeneous:
the epoches of ``typical'' behaviour intermit with the epoches
spent near several least unstable ``untypical'' periodic orbits.

Such a huge inhomogeneity of multipliers evokes the questions:
(i)~How this inhomogeneity is effected by weak noise? (ii)~What
are the consequences for the system dynamics? Indeed, the set of
UPOs is dense and even infinitely small perturbation throws the
system from one UPO onto another. Hence, one can expect that a
very weak noise could first influence the rules of walking over
this set and only then affect properties of UPOs. One can expect
(and the aim of this paper is to validate relevance of such
expectations) the noise to play a ``smoothing'' role, reducing
inhomogeneity of multipliers. As a result of such smoothing, one
should observe, in particular, shrinking of the distribution of
$\langle{z}\rangle$ calculated over finite segments of the chaotic
trajectory because unweighted distribution is more narrow than the
weighted one.

\section{Effect of noise on natural measure distribution}
A significant advance in analytical description of noise action on
systems with fine structure was made for one-dimensional
maps~\cite{Cvitanovic_etal-1998,Dettmann-Howard-2009,Lippolis-Cvitanovic-2010}.
In~\cite{Cvitanovic_etal-1998} the formalism of Feynman diagrams
was employed for treatment of noisy dynamics as a perturbation to
deterministic dynamics over the set of periodic orbits.
In~\cite{Dettmann-Howard-2009} the escape rates for unimodal maps
were evaluated. However, in higher di\-men\-sions---which are
practically inevitable when the map is a Poincar\'e map of chaotic
continuous-time dynamic system---analytical advance becomes
restricted; studies require {\it ad hoc} approaches and/or more
numerics~\cite{Gaspard-2002,Altmann-Endler-2010,Altmann-Leitao-Lopes-2012}.
Studies~\cite{Cvitanovic_etal-1998,Dettmann-Howard-2009,Lippolis-Cvitanovic-2010,Gaspard-2002,Altmann-Leitao-Lopes-2012}
formed solid instrumental and theoretical basis for working with
the noise-induced effects in the systems with fine structure.

Nonetheless, in our study we do not employ the advanced zeta
function formalism and use its limiting version assessing the
measure merely with leading
multipliers~\cite{Cvitanovic_etal-chaosbook}. This version well
proved itself for practical
applications~\cite{Grebogi-Ott-Yorke-1988,Eckhardt-Ott-1994} and
is easier from the view point of calculations when one need as
many UPOs as all the UPOs with $\le20$ loops. Noteworthy, for most
problems addressed in literature, low-length periodic orbits
provide enough data for an accurate
description~\cite{Cvitanovic_etal-1998,Dettmann-Howard-2009},
whereas in our case inhomogeneity of high-length orbits is
essential. Another reason to use simplified (though still
accurate) formalism of~\cite{Grebogi-Ott-Yorke-1988} is that the
analytical theory we will construct is rather qualitative and is
purposed to highlight the mechanism of the effect which will be
finally demonstrated with results of the direct numerical
simulation for chaotic systems.

\begin{figure}[!tb]
\center{
\resizebox{0.45\columnwidth}{!}{%
 \includegraphics
 {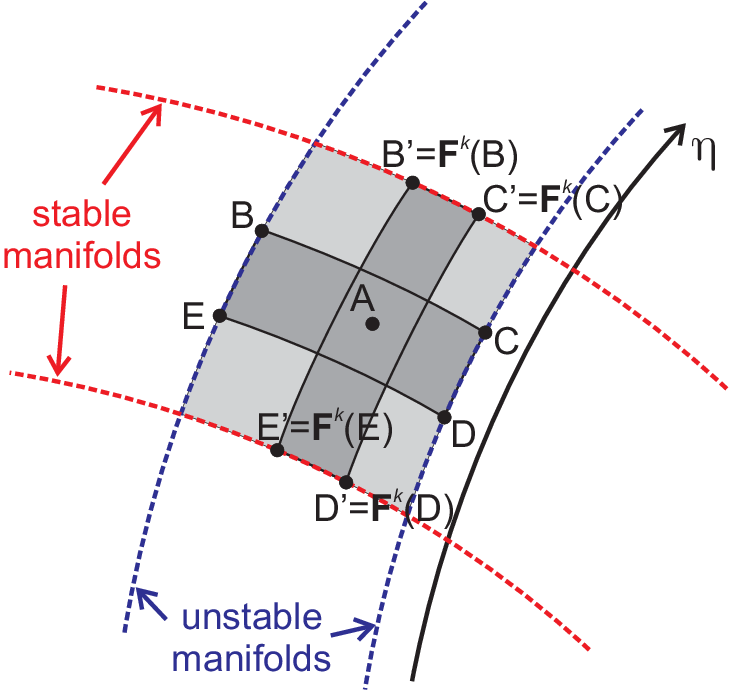} }
}
  \caption{Cell on the Poincar\'e surface
limited by stable and unstable manifolds of the chaotic attractor
in the vicinity of fixed point $\mathrm{A}$. In the vicinity of
fixed point $\mathrm{A}$ of map ${\bf F}^k$, curvilinear
parallelogram $\mathrm{B'C'D'E'}$ is the image of $\mathrm{BCDE}$.
 }
  \label{fig2}
\end{figure}

For evaluation of natural measure, we first briefly recall its
calculation for the no-noise case (interested readers can find a
detailed rigorous consideration in Sec.\,VI
of~\cite{Grebogi-Ott-Yorke-1988}). We consider certain Poincar\'e
section and two-dimensional map on it, which is induced by
three-dimensional phase flow. We are interested in calculation of
the natural measure distribution over attractor on the Poincar\'e
surface. Let us consider $k$ iterations of the map, ${\bf
x}_{n+k}={\bf F}^k({\bf x}_n)$, where $k$ is large, and a small
cell bounded by stable and unstable manifolds, which feature the
directions of compaction and stretching of the phase volume,
respectively, as sketched in Fig.\,\ref{fig2}. In the vicinity of
fixed point $\mathrm{A}$ of map ${\bf F}^k$, curvilinear
parallelogram $\mathrm{B'C'D'E'}$ is formed by states belonging
the cell before $k$ iterations of the map, and $\mathrm{BCDE}$ is
the pre-image of this parallelogram. By definition of the ergodic
(mixing) attractor, for any two subsets $S_a$ and $S_b$ in the
phase space of the system, we have
$$
 \lim_{k\to\infty}\mu[S_a\bigcap{\bf F}^k(S_b)]=\mu(S_a)\,\mu(S_b).
$$
If we take $S_a$ and $S_b$ to be the cell we consider, then
 $\mu(\mbox{cell})=\mu(\mathrm{BCDE})/\mu(\mbox{cell})
 =\mu(\mathrm{B'C'D'E'})/\mu(\mbox{cell})$.
Furthermore, the measure varies smoothly along the unstable
direction~\cite{Bowen-1978} (this is quite general in nature, that
the stretching/expansion is a smoothing process), while it
typically has a complex inhomogeneous structure in the stable
direction, which is transversal to the chaotic attractor.
Therefore, one can easily compare the measure of the cell and
parallelogram $\mathrm{BCDE}$: their ratio is the ratio of their
extension along the unstable direction, because these two sets
have one and the same structure and extension along the
``complex'' inhomogeneous stable direction. We find
$\mu(\mbox{cell})=\mu(\mathrm{BCDE})/\mu(\mbox{cell})
 =\mathrm{CD}/\mathrm{C'D'}=1/\Lambda_{{\bf F}^k}(\mathrm{A})$,
where $\Lambda_{{\bf F}^k}(\mathrm{A})$ is the largest eigen value
of fixed point $\mathrm{A}$ of map ${\bf F}^k$. Recall, that we
have found the measure associated with a single fixed point
$\mathrm{A}$; when there is more fixed points in the cell one has
to sum contributions of all of them,
$$
\mu(\mbox{cell})=\lim_{k\to\infty}\sum_{\mbox{fixed points in
cell}}\frac{1}{\Lambda_{{\bf F}^k}(\mathrm{A}_j)}.
$$
Henceforth, the subscript $j$ indicates the number of a fixed
point or the corresponding UPO. This result was derived
in~\cite{Grebogi-Ott-Yorke-1988}. Recall, the presented evaluation
implies that attractor is hyperbolic, {\it i.e.}, the stable and
unstable manifolds are not mutually tangent anywhere. Although the
Lorenz attractor is non-hyperbolic, it is generally accepted that
statistically significant measure distribution is determined by
the same law $1/\Lambda$.

This calculation of the natural measure sheds light on the fact
that we have to track only the coordinate along the unstable
manifolds, say $\eta$, and evaluate the fraction of states in the
$\varepsilon$-vicinity of some UPO, for which $\eta$ stays within
the $\varepsilon$-vicinity. In no-noise case, the fraction of such
states is simply $1/\Lambda$. Now let us evaluate such a fraction
in the presence of a weak Gaussian $\delta$-correlated noise
$a\xi(t)$: $\langle\xi(t)\rangle=0$,
$\langle\xi(t)\xi(t')\rangle=2\delta(t-t')$, $a$ is the noise
amplitude.

Our map is induced by a continuous time evolution and in the
presence of time-dependent signal, noise, one has to deal with a
continuous-time evolution. Placing the origin of the $\eta$-axis
at fixed point $A_j$, one can find
\begin{equation}
\eta(t_0+T_j)=\Lambda_j\eta(t_0)
 +a\int\limits_0^{T_j}f[{\bf x}_j(\tau)]\,\xi(t_0+\tau)\,e^{\lambda_j(T_j-\tau)}d\tau
 =\Lambda_j\eta(t_0)+\alpha R\,,
\label{eq02}
\end{equation}
where $T_j$ is the period of UPO ${\bf x}_j(t)$ associated with
$A_j$, $f({\bf x})$ is the susceptibility to noise,
$\lambda_j=T_j^{-1}\ln\Lambda_j$ is the leading Lyapunov exponent
of the $j$-th UPO, $R$ is the Gaussian random value of unit
variance, $\langle{R^2}\rangle=1$, and
\begin{eqnarray}
\alpha^2&\equiv &a^2\left\langle\left(\int\limits_0^{T_j}
 f[{\bf x}_j(\tau)]\,\xi(t_0+\tau)\,e^{\lambda_j(T_j-\tau)}d\tau
 \right)^2\right\rangle
\nonumber\\[5pt]
 &=&2a^2\int\limits_0^{T_j}
 f^2[{\bf x}_j(\tau)]\,e^{2\lambda_j(T_j-\tau)}d\tau
 =\frac{a^2F_j^2(\Lambda_j^2-1)}{\lambda_j}\,.
 \nonumber
\end{eqnarray}
Here $F_j$ is the characteristic value of $f[{\bf x}_j(\tau)]$;
\[
 F_j^2\equiv2\lambda_j\,(1-\Lambda_j^{-2})^{-1}
  \int\limits_0^{T_j}f^2[{\bf x}_j(\tau)]\exp(-2\lambda_j\tau)d\tau\,.
\]
According to Eq.\,(\ref{eq02}), the probability of $\eta(t_0+T_j)$
to be at the $\varepsilon_j$-vicinity of fixed point $A_j$ is
\begin{eqnarray}
P(|\eta(t_0+T_j)|\le\varepsilon_j\,|\,\eta(t_0))
 &=&\int\limits_{\frac{-\varepsilon_j-\Lambda_j\eta(t_0)}{\alpha}}^{\frac{\varepsilon_j-\Lambda_j\eta(t_0)}{\alpha}}
 \frac{\displaystyle e^{-\frac{R^2}{2}}}{\sqrt{2\pi}}dR
\nonumber\\
&=&\frac{1}{2}\left[
\Phi\left(\frac{\varepsilon_j-\Lambda_j\eta(t_0)}{\alpha\sqrt{2}}\right)
-\Phi\left(\frac{-\varepsilon_j-\Lambda_j\eta(t_0)}{\alpha\sqrt{2}}\right)
\right],
\nonumber
\end{eqnarray}
where $\Phi(...)$ is the error function. Finally, the measure of
the $\varepsilon_j$-vicinity of fixed point $A_j$, say
$\Omega_{\varepsilon_j}(A_j)$, is
\begin{eqnarray}
\mu[\Omega_{\varepsilon_j}(A_j)]&=&\int\limits_{-\varepsilon_j}^{\varepsilon_j}
  P(|\eta(t_0+T_j)|\le\varepsilon_j\,|\,\eta(t_0))\,d(\eta(t_0))
\nonumber\\
&\approx &\frac{1}{\Lambda_j}\Phi\left(
\frac{\varepsilon_j}{F_ja}\sqrt{\frac{\lambda_j}{2}}
\right).
\label{eq03}
\end{eqnarray}
The last approximate expression corresponds to the limit of large
$\Lambda_j$, which is relevant for long UPOs we consider. For
vanishing noise, $a=0$, the error function turns $1$, and we have
the conventional result of~\cite{Grebogi-Ott-Yorke-1988}.

One can see in Eq.\,(\ref{eq03}), that for larger $\Lambda_j$
(hence, larger $\lambda_j$) the argument of $\Phi$ is typically
larger. Indeed, although $T_j$ also effects the relation between
$\Lambda_j$ and $\lambda_j$, one can find examples of both larger
and smaller $T_j$ for given value of $\lambda_j$. Neglecting
correlation between $T_j$ and $\lambda_j$, one will typically
observe the noticed correlation between $\Lambda_j$ and
$\lambda_j$. This correlation decreases inequivalence between UPOs
with different weight. Maximal decrease of inequivalence is
achieved when the argument of $\Phi$ is small,
$(\varepsilon_j/F_j)\sqrt{\lambda/2}\ll a$\,; at this case one
finds $\mu\propto\sqrt{\lambda_j}/\Lambda_j$.  In particular, for
the $20$-loop UPOs in the Lorenz system the maximal ratio of
weights decreases by 15\%. In Fig.\,\ref{fig1} the distribution of
$\langle{z}\rangle$ with weight $\sqrt{\lambda}/\Lambda$ is
slightly shrunk compared to the one with $\Lambda^{-1}$.

Notice, the vicinity extension $\varepsilon$ is not precisely
known value in our analytical theory. The properties of
neighboring UPOs could be similar and, additionally, the measure
smoothly varies along the unstable direction on the attractor.
Hence, $\varepsilon$ could be not only the characteristic distance
between nearest UPOs of length $N$ but rather a considerably
larger number featuring the size of clusters of UPOs.

\begin{figure}[!t]
\center{
\resizebox{0.60\columnwidth}{!}{%
 \includegraphics
 {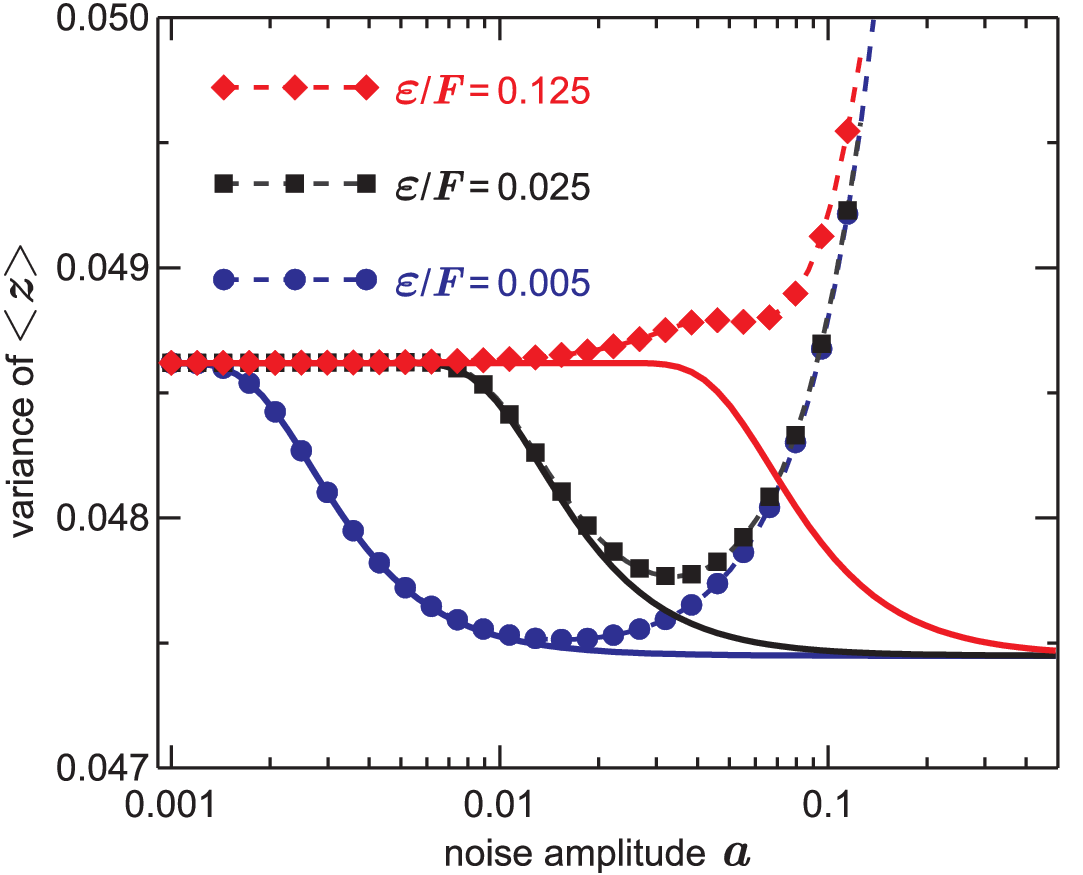} }
}
  \caption{Theoretical dependence of the variance
of $\langle{z}\rangle$ for UPOs of length $N=20$ on noise
amplitude $a$ for the Lorenz system with classical set of
parameters and various values of parameter $\varepsilon$ which
measures characteristic vicinity of UPOs. For the sake of
demonstration we plot the first term of the expression for the
variance [Eq.\,(\ref{eq04})] (solid lines) which features purely
noise-induced redistribution of measure over UPOs of the
noise-free system; the dashed lines with symbols represent the
total variance with account for the noise-induced dispersion of
each single UPO.
 }
  \label{fig3}
\end{figure}

\section{Effect of noise on disorder quantifiers}
Now we calculate the variance of $\langle{z}\rangle$ for UPOs of
length $N$ in the presence of weak noise $a\xi(t)$:
\begin{equation}
\mathrm{var}(\langle{z}\rangle_j)=\sum_j\mu_j\big[\langle{z}\rangle_j^2
 -\overline{\langle{z}\rangle}^2\big]+Ka^2,\quad
 \overline{\langle{z}\rangle}=\sum_j\mu_j\langle{z}\rangle_j.
\label{eq04}
\end{equation}
Here we sum the variance related to the distribution over the set
of UPOs of the noise-free system and the noise-induced distortion
of these UPOs themselves, which is nearly statistically
independent from the former and described by term $Ka^2$. The
value of parameter $K$ is approximately inferred from numerical
calculation of the variance of $\langle{z}\rangle$ for the
segments of the chaotic trajectory (rough analytical estimation
yields the value of the same order of magnitude).

In Fig.\,\ref{fig3}, we plot the variance of $\langle{z}\rangle$
for UPOs of length $N=20$ for the Lorenz system (\ref{eq01}) with
the classical set of parameters. We approximately assume
$\varepsilon_j/F_j$ to be the same for all UPOs (notice, $F_j$ is
of order of $1$). One can note the always-present ordering role of
noise which decreases inequality of weights of UPOs: solid lines
show decrease of the variance by $2.5\%$. However, this ordering
effect can be overwhelmed by the dispersive action of noise on the
orbits when $\varepsilon$ is not small enough.  For the classical
set of parameters the effect is not enough well pronounced and
practically unobservable with dispersion of $\langle{z}\rangle$
for chaotic segments. Still, we report the results for this case,
where the effect is small, for to emphasize their universality:
the effect is always present in some form. For a stronger
inhomogeneity of the attractor one can expect a more pronounced
effect.

\begin{figure}[!t]
\center{
\resizebox{0.60\columnwidth}{!}{%
 \includegraphics
 {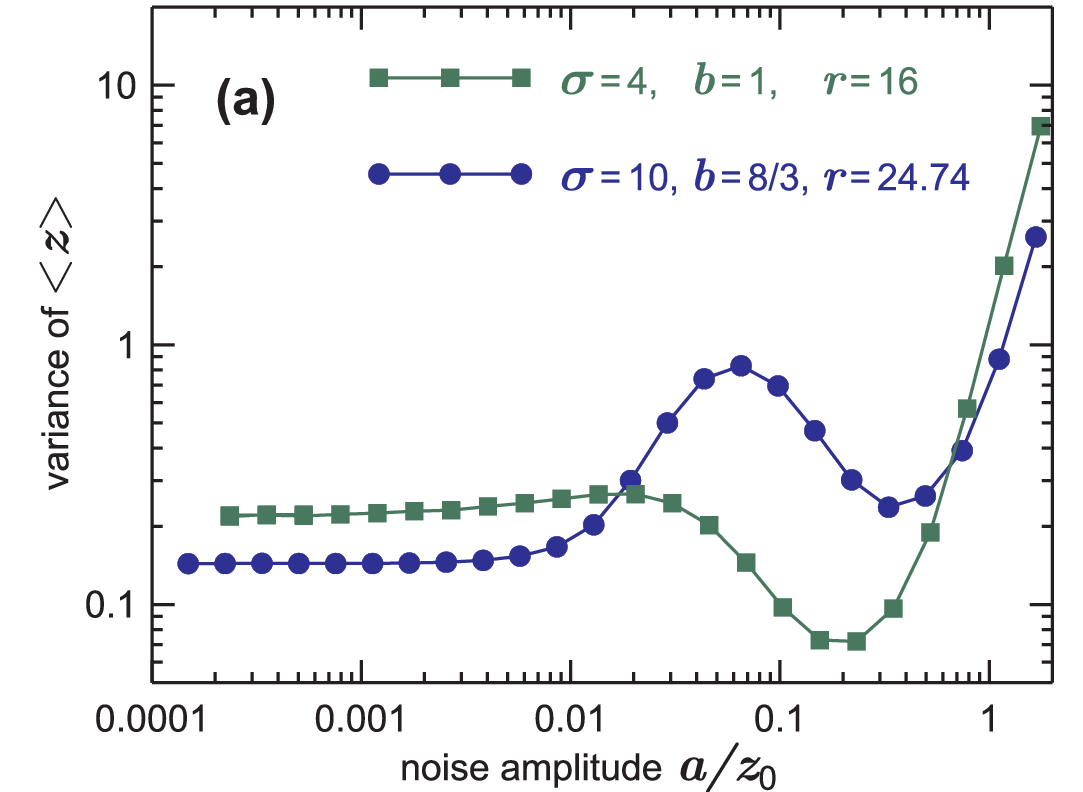} }\\[10pt]
\resizebox{0.60\columnwidth}{!}{%
 \includegraphics
 {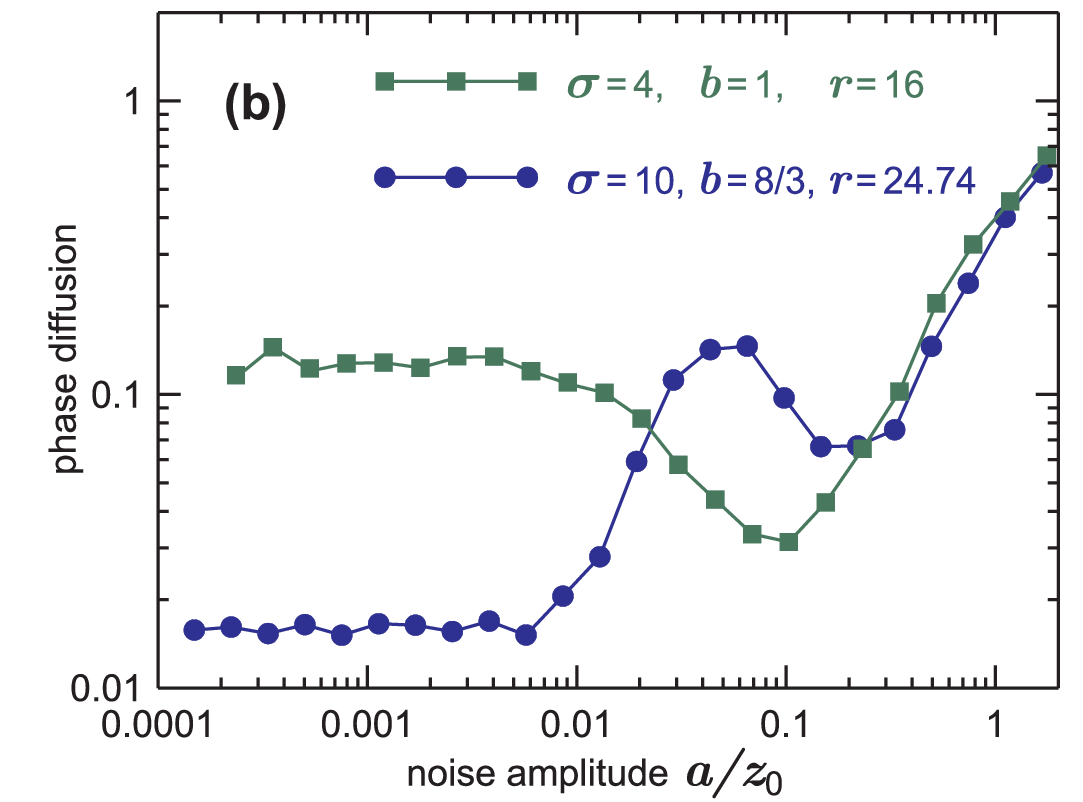} }
}
  \caption{Noise-induced regularization
of chaotic dynamics in the Lorenz system at ($\sigma=1$, $b=8/3$,
$r=24.74$) and ($\sigma=4$, $b=1$, $r=16$): (a) variance of
$\langle{z}\rangle$ calculated over $20$-loop segments of the
chaotic orbit and (b) phase diffusion coefficient measuring
coherence of chaotic oscillations. Noise amplitude $a$ in graphs
is scaled to the coordinate of the saddle points $z_0=r-1$.
 }
  \label{fig4}
\end{figure}

We performed simulations for the Lorenz system (\ref{eq01}) with
white Gaussian noise $a\xi(t)$ put into $dz/dt$. We provide here
the numerical simulation results for additive noise in order to
exclude the trivial deterministic effect of the Stratonovich drift
on the system and show the effect manifestation in its pure form.
However, the analytical theory constructed is valid both for
additive and multiplicative noise. Numerical simulations for
several forms of a multiplicative noise with zero Stratonovich
drift did not reveal anything new compared to the case of additive
noise and we restrict ourselves to presenting the results for the
additive noise only. With the laser interpretation of the Lorenz
system~\cite{Oraevskii-1981}, $z$ measures the inversion of the
population of energetic levels and is both naturally subject to
noise and easily accessible for applying additional additive noise
via noise in the energy pumping.

With the classical values of parameters $\sigma=10$ and $b=8/3$,
the inhomogeneity of the chaotic set is strongest near the
threshold where it becomes attracting, $r\approx24.06$ ({\it
e.g.}, see~\cite{Kaplan-Yorke-1979,Sparrow-1982}). At
$r\approx24.74$ the chaotic attractor becomes the only attractor
in the phase space (between the stability threshold and this
point, it coexists with the stable fixed points which
significantly influence the noise-perturbed dynamics of the
system). In Fig.\,\ref{fig4}(a), the dependence of
$\mathrm{var}\langle{z}\rangle$ for $20$-loop segments of the
chaotic trajectory on the noise amplitude exhibits a
well-pronounced minimum and resembles the theoretical dependencies
in Fig.\,\ref{fig3}. Indeed, the dependencies in Fig.\,\ref{fig4}a
may be decomposed into superposition of a stepwise drop (compare
to the solid lines in Fig.\,\ref{fig3}) and quadratic growth of
$\mathrm{var}\langle{z_j}\rangle$ due to noisy distortion of
trajectories; exactly such superpositions form theoretical
dependencies plotted in Fig.\,\ref{fig3}. The disorder-reduction
effect for some level of noise is well pronounced though this
local minimum corresponds to larger dispersion than in the
noise-free case (circle-marked blue curve in the plot). In
reality, the level of noise cannot be zero and this local minimum
can provide a smaller dispersion than that for the minimal noise
level if, for instance, noise cannot be diminished below
$a=0.05\,z_0$ ($z_0=r-1$ is the coordinate of the saddle points).
Furthermore, the local minimum can become a global one for a
larger inhomogeneity of the attractor. In Fig.\,\ref{fig4}(a), the
dependence of $\mathrm{var}\langle{z}\rangle$ on the noise
amplitude for $\sigma=4$, $b=1$, $r=16$ (where the chaotic
attractor nearly touches the slightly unstable non-trivial saddle
points) has the minimum which provides 3 times smaller dispersion
than the noise-free case.

The dispersion of $\langle{z}\rangle$ calculated over finite
segments of the chaotic trajectory is only a sample of a
quantifier with which one can observe the ordering effect of noise
we report. This ordering can be observed with other quantifiers of
the system dynamics as well. For instance, one of the very
important characteristics of the chaotic systems is the coherence
of their oscillations. In particular, coherence determines
susceptibility of the system to control forcing and predisposition
to synchronization ({\it e.g.},
see~\cite{Pikovsky-Rosenblum-Kurths-2001,Goldobin-Rosenblum-Pikovsky-2003,Goldobin-2008}).
The coherence is quantified by the diffusion coefficient of the
chaotic oscillation phase (the oscillation phase to be introduced
so that it growths by $2\pi$ for one revolution of the
trajectory). In Fig.\,\ref{fig4}(b) the dependencies of the phase
diffusion coefficient on the noise amplitude are plotted for the
same sets of parameters as we used for calculation of the
dispersion of $\langle{z}\rangle$. One can see well pronounced
minima. For $\sigma=4$, $b=1$, $r=16$ the diffusion can be
suppressed by factor $5$ in comparison with the noise-free case.

\begin{figure}[!t]
\center{
\resizebox{0.60\columnwidth}{!}{%
 \includegraphics
 {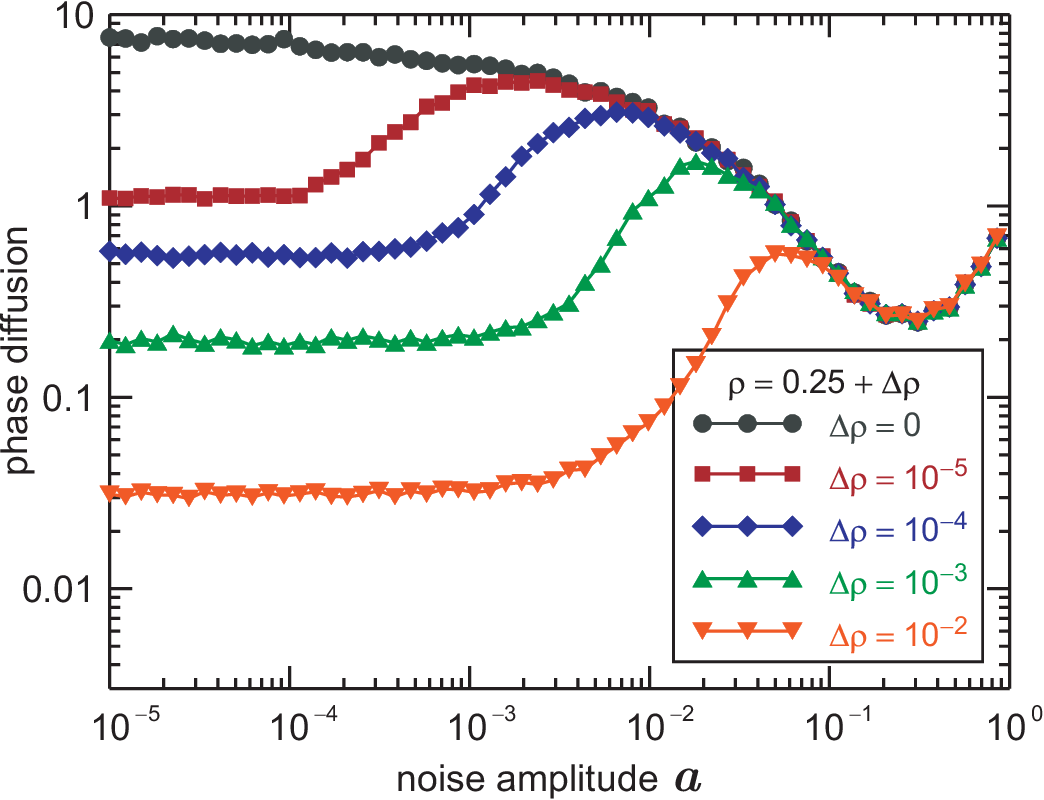} }
}
  \caption{Noise-induced regularization
of chaotic dynamics for mapping (\ref{eq05}).
 }
  \label{fig5}
\end{figure}

\section{Observation of the noise-induced regularization effect with mapping}
Our analytical treatment is not restricted to the Lorenz system
which is utilized only for demonstration.  The Lorenz system is
known to possess significant peculiarities~\cite{Sparrow-1982};
therefore it would be appropriate to validate the effect we
observe with independent examples. The classical scenario of the
transition to chaos in the Lorenz system is known to be a special
case of the transition to chaos via cascade of bifurcations of
homoclinic loops~\cite{Lyubimov-Zaks-1983}. Within the vicinity of
the bifurcation accumulation point, a general system, where such a
scenario of the transition to chaos occurs, can be described by
the following mapping (in the noise-free
case)~\cite{Lyubimov-Zaks-1983};
\begin{eqnarray}
x_{n+1}&=&\left\{
\begin{array}{cc}
x_n^\nu-\rho_1\,,&\mbox{ for }x_n>0\,;\\ [5pt]
-A_2|x_n|^\nu+\rho_2\,,&\mbox{ for }x_n<0\,;
\end{array}
\right.
\nonumber\\ [5pt]
t_{n+1}&=&t_n+1+\left\{
\begin{array}{cc}
T_1\ln(\rho_1/x_n)\,,&\mbox{ for }x_n>0\,;\\ [5pt]
T_2\ln(\rho_2/|x_n|)\,,&\mbox{ for }x_n<0\,;
\end{array}
\right.
\nonumber
\end{eqnarray}
where $x_n$ is the system state at the time moment $t_n$, $\nu$ is
the saddle index, $\rho_{i}$ are the parameters featuring
deviation from the bifurcation accumulation point
($\rho_1=\rho_2=0$ at this point). The iteration $n\to n+1$
corresponds to the $(n+1)$-th revolution of the system---the
oscillation phase grows for $2\pi$ as $n\to n+1$. For the sake of
definiteness we restrict ourselves to the case of
$\rho_1=\rho_2\equiv\rho$, $\nu=1/2$, $A_2=1$, $T_1=T_2=1$ and
introduce noise:
\begin{equation}
\begin{array}{c}
\displaystyle
x_{n+1}=\mathrm{signum}(x_n)(\sqrt{|x_n|}-\rho)+a\zeta_n\,;\\[5pt]
 t_{n+1}=t_n+1+\ln(\rho/|x_n|).
\end{array}
\label{eq05}
\end{equation}
Here $a$ is the noise amplitude and $\zeta_n$ are independent
random numbers uniformly distributed in
$\left[-\sqrt{3},\sqrt{3}\right]$ ({\it i.e.},
$\langle\zeta^2\rangle=1$).

For this mapping the phase grows uniformly with each iteration
while time increments $(t_{n+1}-t_n)$ are non-constant.  The phase
diffusion can be calculated from $t_n$ as a function of $n$ (see,
{\it e.g.},~\cite{Reimann-etal-2001}). In Fig.\,\ref{fig5} one can
see that the regularizing effect of noise can be as well observed
near $\rho=0.25$ where the natural measure distribution is highly
inhomogeneous.

\section{Conclusion}
We have reported the ordering effect of noise on the chaotic
dynamics. This effect is rooted into inhomogeneity of the natural
measure over the attractor and the fact that noise diminishes this
inhomogeneity. Without noise, the measure is determined by the
largest multipliers of unstable periodic orbits (UPOs) composing
the chaotic attractor and is typically extremely nonuniform for
long orbits due to the exponential dependence of the multiplier on
the orbit length and the Lyapunov exponent which vary for
different UPOs. Generally, the unweighted probability density
function of some average for UPOs is centered around the value
which does not have to correspond to the maximal measure (see
Fig.\,\ref{fig1} where the big tail of the weighted PDF clearly
indicates that the natural measure of UPOs with large
$\langle{z}\rangle$ is drastically larger than that of UPOs near
the peak of the unweighted PDF). Hence, the dynamics is more
contributed by UPOs which are not ``geometrically typical''. Noise
smooths the measure over the set of UPOs and reduces the role of
excursions along the ``untypical'' UPOs. On these grounds, we
expect the effect to be a universal feature for chaotic dynamics
with exception for some model systems with absolutely uniform
natural measure (like the tent map). With strong enough
inhomogeneity given, the ordering effect can overwhelm the
distorting action of noise on UPOs and lead to significant
observable ordering of the system dynamics. We have developed the
analytical theory for the effect of noise on the measure
distribution over chaotic attractor and observed the ordering
effects for dispersion of averages over finite segments of the
chaotic trajectory and coherence of chaotic oscillations.
Noteworthy, the reported phenomenon is novel; its mechanism being
robust and vital is not related to any of well-known phenomena of
noise-induced ordering
(stochastic~\cite{McNamara-Wiesenfeld-1989,Gammaitoni-etal-1998}
or coherence resonance~\cite{Pikovsky-Kurths-1997}, suppression of
deterministic chaos by noise~\cite{Anishchenko-1995}, {\it etc.}).

The particular example of potential practical application of the
phenomenon discussed is the control of single-mode semiconductor
and gas lasers, where increase of the energy pumping (and lasing
power) leads to chaotic radio-frequency self-modulation of the
amplitude of the laser emission
wave~\cite{Oraevskii-1981,Boccaletti-Allaria-Meucci-2004}. Such a
modulation of signal might be undesirable as it leads to
broadening of the power spectrum and, therefore, difficulties for
focusing the laser beam or redirecting it without angular
divergence. The decrease of coherence of the amplitude or phase
modulation as well impairs opportunities for information transfer
and employability of the system for fiber-optics communication.
Synchronization phenomena and the coherence property for laser
systems also provide opportunities related to cryptography ({\it
e.g.}, see~\cite{Banerjee-etal-2011a,Banerjee-etal-2011b}). In
general, the coherence determines the precision of clocks
(including biological
ones~\cite{Montminy-1997,Bell-Pedersen-etal-2005}), the quality of
electric generators, the susceptibility of oscillatory systems to
external driving ({\it e.g.},
see~\cite{Goldobin-Rosenblum-Pikovsky-2003,Goldobin-2008}) and,
therefore, their predisposition to synchronization.

\acknowledgement{
The author thanks Michael Zaks and Alexander Gorban for fruitful
comments on the work.
 }


\end{document}